\documentclass[12pt]{iopart}

\usepackage[dvips]{graphicx}

\begin{document}

\title[Mapping Koch curves into scale-free small-world networks]{Mapping Koch curves into scale-free small-world networks}

\author{Zhongzhi Zhang$^{1,2}$, Shuyang Gao$^{1,2}$, Lichao Chen$^{3}$, Shuigeng Zhou$^{1,2}$, Hongjuan Zhang$^{2,4}$ and  Jihong Guan$^{5}$}
\address{$^1$School of Computer Science, Fudan
University, Shanghai 200433, China\\
$^2$Shanghai Key Lab of Intelligent Information Processing, Fudan
University, Shanghai 200433, China\\
$^{3}$Electrical Engineering Department, University of California,
Los Angeles, CA 90024, USA\\
$^4$Department of Mathematics, College of Science, Shanghai
University, Shanghai
200444, China\\
$^5$Department of Computer Science and Technology, Tongji
University, 4800 Cao'an Road, Shanghai 201804, China}

\ead{zhangzz@fudan.edu.cn,sgzhou@fudan.edu.cn,jhguan@tongji.edu.cn}

\begin{abstract}
The class of Koch fractals is one of the most interesting families
of fractals, and the study of complex networks is a central issue in
the scientific community. In this paper, inspired by the famous Koch
fractals, we propose a mapping technique converting Koch fractals
into a family of deterministic networks, called Koch networks. This
novel class of networks incorporates some key properties
characterizing a majority of real-life networked systems---a
power-law distribution with exponent in the range between 2 and 3, a
high clustering coefficient, small diameter and average path length,
and degree correlations. Besides, we enumerate the exact numbers of
spanning trees, spanning forests, and connected spanning subgraphs
in the networks.  All these features are obtained exactly according
to the proposed generation algorithm of the networks considered. The
network representation approach could be used to investigate the
complexity of some real-world systems from the perspective of
complex networks.
\end{abstract}



\maketitle


\section{Introduction}

The past decade has witnessed a great deal of activity devoted to
complex networks by the scientific community, since many systems in
the real world can be described and characterized by complex
networks~\cite{AlBa02,DoMe02,Ne03,BoLaMoChHw06}. Prompted by the
computerization of data acquisition and the increased computing
power of computers, researchers have done a lot of empirical studies
on diverse real networked systems, unveiling the presence of some
generic properties of various natural and manmade networks:
power-law degree distribution $P(k) \sim k^{-\gamma}$ with
characteristic exponent $\gamma$ in the range between 2 and
3~\cite{BaAl99}, small-world effect including large clustering
coefficient and small average distance~\cite{WaSt98}, and degree
correlations~\cite{Newman02,Newman03c}.

The empirical studies have inspired researchers to construct network
models with the aim to reproduce or explain the striking common
features of real-life systems~\cite{AlBa02,DoMe02}. In addition to
the seminal Watts-Strogatz's (WS) small-world network
model~\cite{WaSt98} and Barab\'asi-Albert's (BA) scale-free network
model~\cite{BaAl99}, a considerable number of models and mechanisms
have been developed to mimic real-world systems, including initial
attractiveness~\cite{DoMeSa00}, aging and cost~\cite{AmScBaSt00},
fitness model~\cite{BiBa01}, duplication~\cite{ChLuDeGa03}, weight
or traffic driven evolution~\cite{BaBaVe04a,WaWaHuYaOu05},
geographical constraint~\cite{RoCoAvHa02}, accelerating
growth~\cite{MaGa05,ZhFaZhGu09}, coevolution~\cite{GrBl08},
visibility graph~\cite{LaLuBaLuNu08}, to list a few. Although
significant progress has been made in the field of network modeling
and has led to a significant improvement in our understanding of
complex systems, it is still a fundamental task and of current
interest to construct models mimicking real networks and reproducing
their generic properties from different angles~\cite{PaLoVi10}.

In this paper, enlightened by the famous class of Koch fractals, we
propose a family of deterministic mathematical networks, called Koch
networks, which integrates the observed properties of real networks
in a single framework. We derive analytically exact scaling laws for
degree distribution, clustering coefficient, diameter, average
distance or average path length (APL), degree correlations, even for
spanning trees, spanning forests, and connected spanning subgraphs.
The obtained precise results show that Koch networks have rich
topological features: they obey power-law degree distribution with
exponent lies between 2 and 3; they have a large clustering
coefficient, their diameter and APL grow logarithmically with the
total number of nodes; and they may be either disassortative or
uncorrelated.

This work unfolds an alternative perspective in the study of complex
networks. Instead of searching generation mechanisms for real
networks, we explore deterministic mathematical networks that
exhibit some typical properties of real-world systems. As the
classical Koch fractals are important for the understanding of
geometrical fractals in real systems~\cite{Ma82}, we believe that
Koch networks could provide valuable insights into real-world
systems.

\begin{figure}
\begin{center}
\includegraphics[width=.5\linewidth,trim=100 0 100 0]{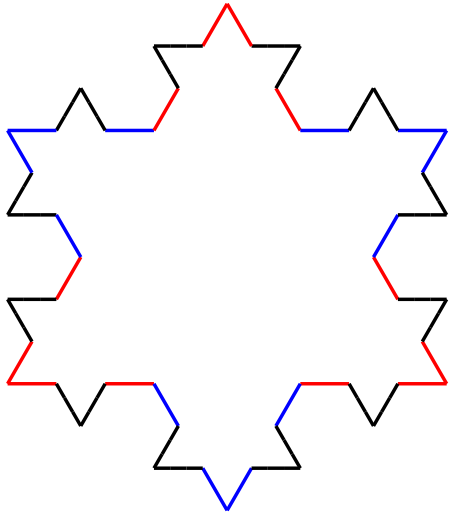} \
\end{center}
\caption[kurzform]{\label{curve1} (Color online) The first two
generations of the construction for Koch curve.}
\end{figure}

\section{Network construction}

In order to define the networks, we first introduce a classical
fractal---Koch curve, which was proposed by Helge von
Koch~\cite{Ko1906}. The Koch curve, denoted as $S_1(t)$ after $t$
generations, can be constructed in a recursive way. To produce this
well-known fractal, we begin with an equilateral triangle and let
this initial configuration be $S_1(0)$. In the first generation, we
perform the following operations: firstly, we trisect each side of
the initial equilateral triangle; secondly, on the middle segment of
each side, we construct new equilateral triangles whose interiors
lie external to the region exclosed by the base triangle; thirdly,
we remove the three middle segments of the base triangle, upon which
new triangles were established. Thus, we get $S_1(1)$. In the second
generation, for each line segment in $S_1(1)$, repeat above
procedure of three operations to obtain $S_1(2)$. This process is
then repeated for successive generations. As $t$ tends to infinite,
the Koch curve is obtained, and its Hausdorff dimension is
$d_f=\frac{2\ln 2}{\ln3}$~\cite{LaVaMeVa87}. Figure~\ref{curve1}
depicts the structure of $S_1(2)$.

\begin{figure}
\begin{center}
\includegraphics[width=0.5\linewidth,trim=100 0 100 0]{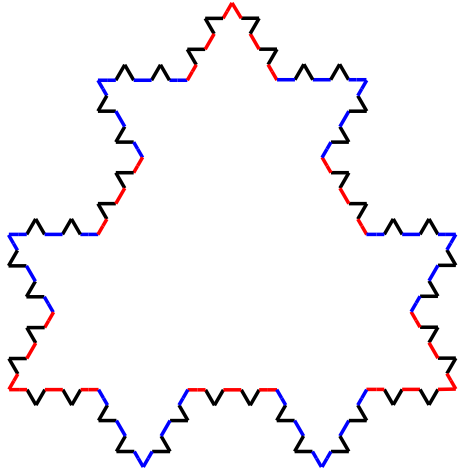} \
\end{center}
\caption[kurzform]{\label{curve2} (Color online) The first two
generations of the construction for generalized Koch curve in the
case of $m=2$.}
\end{figure}

Koch curve can be easily generalized to other dimensions by
introducing a parameter $m$ (a positive
interger)~\cite{LaVaMeVa87,Sc65}. Denoted by $S_m(t)$ the
generalization after $t$ generations, which is constructed as
follows~\cite {LaVaMeVa87}: Start with an equilateral triangle as
the initial configuration $S_m(0)$. In the first generation, we
perform the following operations similar to those described in last
paragraph: partition each side of the initial triangle into $2m+1$
segments, which are consecutively numbered $1, 2,\cdots, 2m, 2m+1$
from one endpoint of the side to the other; construct a new small
equilateral triangle on each even-numbered segment so that the
interiors of the new triangles lie in the exterior of the base
triangle; remove the segments upon which triangles were constructed.
In this way we obtain $S_m(1)$. Analogously, we can get $S_m(t)$
from $S_m(t-1)$ by repeating recursively the procedure of above
three operations for each existing line segments in generation
$t-1$. In the infinite $t$ limit, the Hausdorff dimension of the
generalized Koch curves $d_f=\frac{\ln
(4m+1)}{\ln(2m+1)}$~\cite{LaVaMeVa87}. Figure~\ref{curve2} shows the
structure of $S_2(2)$.

\begin{figure}
\begin{center}
\includegraphics[width=0.6\linewidth,trim=100 0 100 0]{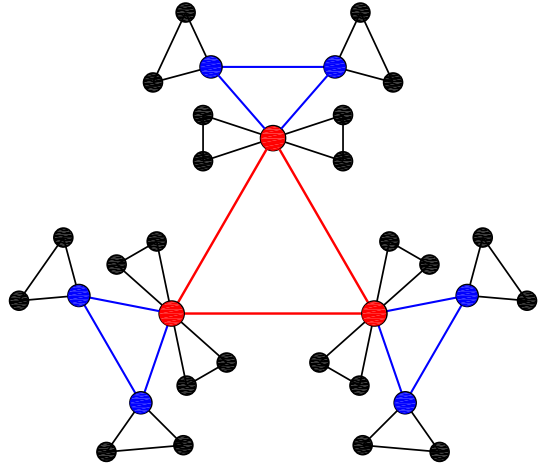}
\end{center}
\caption[kurzform]{\label{network1} (Color online) The network
derived from $S_1(2)$.}
\end{figure}

\begin{figure}
\begin{center}
\includegraphics[width=0.6\linewidth,trim=100 0 100 0]{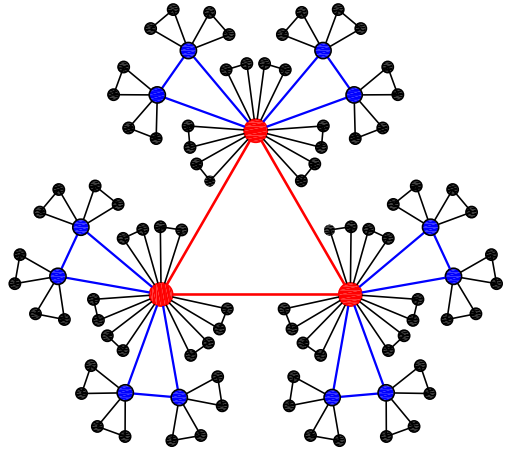}
\end{center}
\caption[kurzform]{\label{network2}(Color online) The network
corresponding to $S_2(2)$.}
\end{figure}

The generalized Koch curves can be used as a basis of a new class of
networks: sides (excluding those deleted) of the triangles of the
Koch curves constructed at arbitrary generations are mapped to
nodes, which are connected to one another if their corresponding
sides in the Koch curves are in contact. For uniformity, the three
sides of the initial equilateral triangle of $S_m(0)$ also
correspond to three different nodes. We shall call the resultant
networks Koch networks. Note that after the birth of each side of a
triangle constructed at a given generation of the Koch curves,
although some segments of it will be removed at subsequent steps, we
look on its remain segments as a whole and map it to only one node.
Figures~\ref{network1} and~\ref{network2} show two networks
corresponding to $S_1(2)$ and $S_2(2)$, respectively.

Obviously, Koch networks have an infinite number of nodes. But in
what follows we shall generally consider the network characteristics
after a finite number of generations in the development of the
complete Koch networks. From our analytical results, we can quickly
obtain the characteristics of the complete networks by taking the
limit of large $t$. However, the numerical results are necessarily
limited to networks with finite order (number of all nodes).

\section{Generation algorithm}

\begin{figure}
\begin{center}
\includegraphics[width=0.50\linewidth,trim=100 0 100 0]{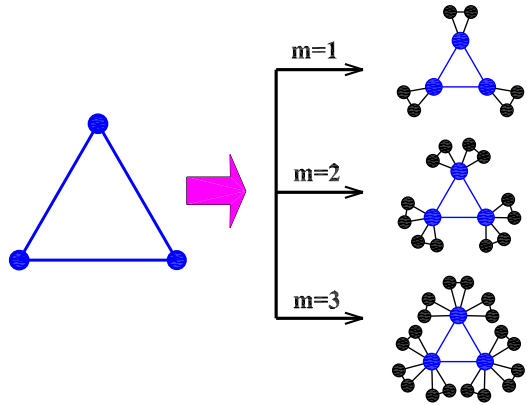}
\end{center}
\caption[kurzform]{\label{iterative} (Color online) Iterative
construction method for the Koch networks. }
\end{figure}

According to the construction process of the generalized Koch curves
and the proposed method of mapping from Koch curves to Koch
networks, we can introduce with ease an iterative algorithm to
create Koch networks, denoted by $K_{m,t}$ after $t$ generation
evolutions. The algorithm is as follows: Initially ($t=0$),
$K_{m,0}$ consists of three nodes forming a triangle. For $t\geq 1$,
$K_{m,t}$ is obtained from $K_{m,t-1}$ by adding $m$ groups of nodes
for each of the three nodes of every existing triangles in
$K_{m,t-1}$. Each node group has two nodes. These two new nodes and
its ``mother'' node are linked to one another shaping a new
triangle. In other word, to obtain $K_{m,t}$ from $K_{m,t-1}$, we
replace each of the existing triangles of $K_{m,t-1}$ by the
connected clusters on the rightmost side of Fig.~\ref{iterative}.
Figures~\ref{network1} and~\ref{network2} illustrate the growing
process of the networks for two particular cases of $m=1$ and $m=2$,
respectively. Notice that in the peculiar case of $m=1$, the
networks under consideration reduce to the one previously studied
in~\cite{ZhZhXiChGu09}.

Let us compute the order and size (number of all edges) of Koch
networks $K_{m,t}$. To this end, we first consider the total number
of triangles $L_\Delta(t)$ that exist at step $t$. By construction
(see Fig.~\ref{iterative}), this quantity increases by a factor of
$3m+1$, i.e., $L_\Delta(t)=(3m+1)\,L_\Delta(t-1)$. Considering
$L_\Delta(0)=1$, we have $L_\Delta(t)=(3m+1)^t$. Denote $L_v(t)$ and
$L_e(t)$ as the numbers of nodes and edges created at step $t$,
respectively. Note that each triangle in $K_{m,t-1}$ will give birth
to $6m$ new nodes and $9m$ new edges at step $t$, then one can
easily obtain: $L_v(t)=6m\,L_\Delta(t-1)=6m\,(3m+1)^{t-1}$ and
$L_e(t)=9m\,L_\Delta(t-1)=9m\,(3m+1)^{t-1}$, both of which hold for
arbitrary $t>0$. Then, the total numbers of nodes $N_t$ and edges
$E_t$ present at step $t$ are
\begin{equation}\label{Nt}
N_t=\sum_{t_i=0}^{t}L_v(t_i)=2\,(3m+1)^{t}+1
\end{equation}
and
\begin{equation}\label{Et}
E_t =\sum_{t_i=0}^{t}L_e(t_i)=3\,(3m+1)^{t},
\end{equation}
respectively. Thus, the average degree is
\begin{equation}\label{AveDeg}
\langle k \rangle =\frac{2\,E_t}{N_t} =
\frac{6\,(3m+1)^{t}}{2\,(3m+1)^{t}+1},
\end{equation}
which is approximately $3$ for large $t$, showing that Koch networks
are sparse as most real-life
networks~\cite{AlBa02,DoMe02,Ne03,BoLaMoChHw06}.

\section{Topological properties}

Now we study some relevant characteristics of the Koch networks
$K_{m,t}$, focusing on degree distribution, clustering coefficient,
diameter, average distance, degree correlations, spanning trees,
spanning forests, and connected spanning subgraphs. We emphasize
that this is the first analytical study for counting spanning trees,
spanning forests, and connected spanning subgraphs in scale-free
networks.

\subsection{Degree distribution}

Let $k_i(t)$ be the degree of a node $i$ at time $t$. When node $i$
enters the network at step $t_i$ ($t_i\geq 0$), it has a degree of
$2$, viz. $k_i(t_i)=2$. To determine $k_i(t)$, we first consider the
number of triangles involving node $i$ at step $t$ that is denoted
by $L_\Delta(i,t)$. These triangles will give rise to new nodes
linked to the node $i$ at step $t+1$. Then at step $t_i$,
$L_\Delta(i, t_i)=1$. By construction, for any triangle involving
node $i$ at a given step, it will lead to $m$ new triangles passing
by node $i$ at next step. Thus,
$L_\Delta(i,t)=(m+1)\,L_\Delta(i,t-1)$. Considering the initial
condition $L_\Delta(i, t_i)=1$, we have
$L_\Delta(i,t)=(m+1)^{t-t_{i}}$. On the other hand, every triangle
passing by node $i$ contains two links connected to $i$, therefore
we have $k_i(t)=2\,L_\Delta(i,t)$. Then we obtain
\begin{equation}\label{ki}
k_i(t)=2\,L_\Delta(i,t)=2(m+1)^{t-t_{i}}.
\end{equation}
In this way, at time $t$ the degree of arbitrary node $i$ of Koch
networks has been computed explicitly. From Eq.~(\ref{ki}), it is
easy to see that at each step the degree of a node increases $m$
times, i.e.,
\begin{equation}\label{ki2}
k_i(t)=(m+1)\,k_i(t-1).
\end{equation}

Equation~(\ref{ki}) shows that the degree spectrum of Koch networks
is discrete. Thus, we can get the degree distribution $P(k)$ of the
Koch networks via the cumulative degree distribution~\cite{Ne03}
given by
\begin{equation}\label{pcumk}
P_{\rm cum}(k)=\frac{1}{N_t}\,\sum_{\tau \leq t_i}L_v(\tau)
={2\times (3m+1)^{t_i}+1 \over 2\times (3m+1)^{t}+1}.
\end{equation}
Substituting for $t_i$ in this expression using $t_i=t-\frac{\ln
(\frac{k}{2})}{\ln (m+1)}$ gives
\begin{equation}
P_{\rm cum}(k)=\frac{2\times (3m+1)^{t}\times
\left(\frac{k}{2}\right)^{-\frac{\ln(3m+1)}{\ln(m+1)}}+1} {2\times
(3m+1)^{t}+1}.
\end{equation}
In the infinite $t$ limit, we obtain
\begin{equation}\label{gammak}
P_{\rm cum}(k)=2^{\frac{\ln(3m+1)}{\ln(m+1)}} \times
k^{-\frac{\ln(3m+1)}{\ln(m+1)}}.
\end{equation}
So the degree distribution follows a power-law form $P(k)\sim
k^{-\gamma}$ with the exponent $\gamma=1+\frac{\ln(3m+1)}{\ln(m+1)}$
belonging to the interval $[2,3]$. When $m$ increases from 1 to
infinite, $\gamma$ decreases from 3 to 2. It should be stressed that
the exponent of degree distribution of most real scale-free networks
also lies in the same range between 2 and 3.

\subsection{Clustering coefficient}

By definition, the clustering coefficient~\cite{WaSt98} of a node
$i$ with degree $k_i$ is the ratio between the number of triangles
$e_i$ that actually exist among the $k_i$ neighbors of node $i$ and
the maximum possible number of triangles involving $i$,
$k_i(k_i-1)/2$, namely, $C_i =2e_i/[k_i(k_i-1)]$. For Koch networks,
we can obtain the  exact expression of clustering coefficient $C(k)$
for a single node with degree $k$. By construction, for any given
node having a degree $k$, there are just $e=\frac{k}{2}$ triangles
connected this node, see also Eq. (\ref{ki}). Hence there is a
one-to-one corresponding relation between the clustering coefficient
of a node and its degree: for a node of degree $k$,
\begin{equation}\label{Ck}
C(k)=\frac{1}{k-1},
\end{equation}
which shows a power-law scaling $C(k)\sim k^{-1}$ in the large limit
of $k$, in agreement with the behavior observed in a variety of
real-life systems~\cite{RaBa03}.

\begin{figure}
\begin{center}
\includegraphics[width=0.25\linewidth,trim=110 10 120 20]{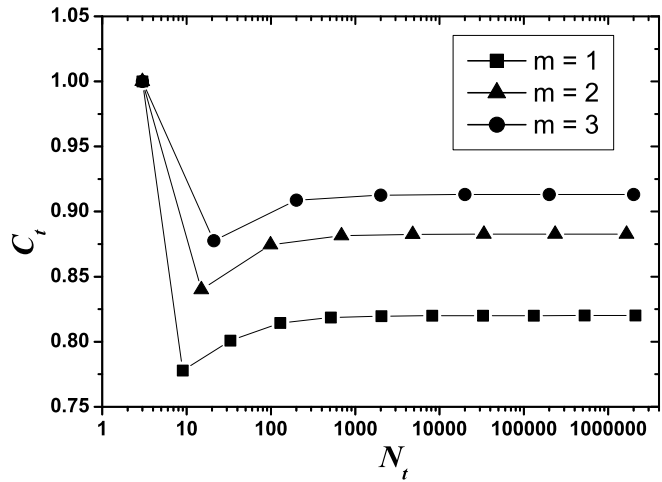} \
\end{center}
\caption[kurzform]{\label{clustering} Semilogarithmic plot of
average clustering coefficient $C_t$ versus network order $N_{t}$.}
\end{figure}

After $t$ step growth, the average clustering coefficient $C_t$ of
the whole network $K_{m,t}$, defined as the mean of $C_i^{'}s$ over
all nodes in the network, is given by
\begin{equation}\label{ACC}
C_t=\frac{1}{N_{t}}\sum_{r=0}^{t}
    \left [\frac{1}{G_r-1}\times L_v(r)\right],
\end{equation}
where the sum runs over all the nodes of all generations and $G_r$
is the degree of those nodes created at step $r$, which is given by
Eq.~(\ref{ki}). In the limit of large $N_{t}$, equation~(\ref{ACC})
converges to a nonzero value $C$, as reported in
Fig.~\ref{clustering}. For $m=1$, 2, and 3, $C$ are 0.82008,
0.88271, and 0.91316, respectively. As $m$ approaches infinite, $C$
converges to 1. Thus, $C$ increases with $m$: when $m$ grows from 1
to infinite, $C$ increases form 0.82008 to 1. Therefore, for the
full range of $m$, the the average clustering coefficient of Koch
networks is very high.

\subsection{Diameter}

Most real networks are small-world, i.e., their average distance
grows logarithmically with network order or slower. Here average
distance means the minimum number of edges connecting a pair of
nodes, averaged over all node pairs. For a general network, it is
not easy to derive a closed formula for its average distance.
However, the whole family of Koch networks has a self-similar
structure, allowing for analytically calculating the average
distance, which approximately increases as a logarithmic function of
network order. We leave the detailed exact derivation about average
distance to next subsection.

Here we provide the exact result of the diameter of $K_{m,t}$
denoted by $Diam(K_{m,t})$ for all parameter $m$, which is defined
as the maximum of the shortest distances between all pairs of nodes.
Small diameter is consistent with the concept of small-world. The
obtained diameter also scales logarithmically with the network
order. The computation details are presented as follows.

Clearly, at step $t=0$, $Diam(K_{m,0})$ is equal to 1. At each step
$t\geq 1$, we call newly-created nodes at this step \emph{active
nodes}. Since all active nodes are attached to those nodes existing
in $K_{m,t-1}$, so one can easily see that the maximum distance
between any active node and those nodes in $K_{m,t-1}$ is not more
than $Diam(K_{m,t-1})+1$ and that the maximum distance between any
pair of active nodes is at most $Diam(K_{m,t-1})+2$. Thus, at any
step, the diameter of the network increases by 2 at most. Then we
get $2(t+1)$ as the diameter of $Diam(K_{m,t})$. Equation~(\ref{Nt})
indicates that the logarithm of the order of $Diam(K_{m,t})$ is
proportional to $t$ in the large limit $t$. Thus the diameter
$Diam(K_{m,t})$ grows logarithmically with the network order,
showing that the Koch networks are small-world.

\subsection{Average path length}

Using a method similar to but different from those in
literature~\cite{HiBe06,ZhChZhFaGuZo08}, we now study analytically
the average path length $d_t$ of Koch networks $K_{m,t}$. It follows
that
\begin{equation}\label{eq:app1}
d_{t} = \frac{D_{\rm tot}(t)}{N_t(N_t-1)/2}\,,
\end{equation}
where $D_{\rm tot}(t)$ is the total distance between all couples of
nodes, i.e.,
\begin{equation}\label{eq:app2}
D_{\rm tot}(t) = \sum_{i \in K_{m,t},\, j \in K_{m,t},\, i \neq j}
d_{ij}(t),
\end{equation}
in which $d_{ij}(t)$ is the shortest distance between node $i$ and
$j$ in networks $K_{m,t}$.

\begin{figure}
\begin{center}
\includegraphics[width=0.7\linewidth,trim=100 0 100 0]{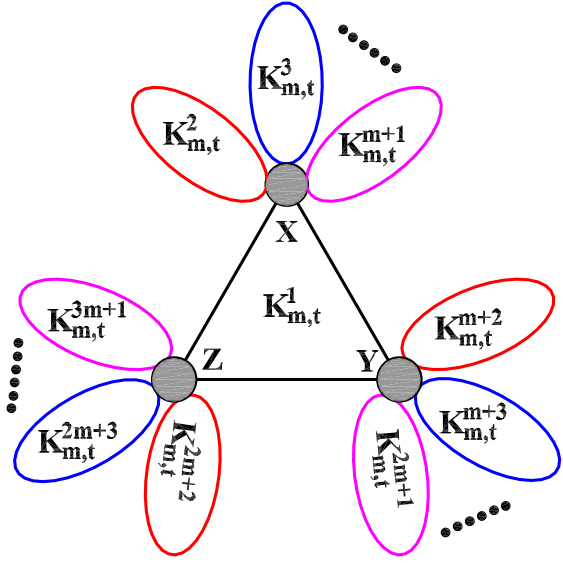}
\caption{(Color online) Second construction method of Koch networks
that highlights self-similarity. The graph after $t+1$ construction
steps, $K_{m,t+1}$, consists of $3m+1$ copies of $K_{m,t}$ denoted
as $K_{m,t}^{\theta}$ $(\theta=1,2,3,\ldots,3m, 3m+1)$, which are
connected to one another as above.} \label{copy}
\end{center}
\end{figure}

Notice that Koch networks have a self-similar structure, which
allows us to address $D_{\rm tot}(t)$ analytically. This
self-similar structure is obvious from an equivalent network
construction method: to obtain $K_{m,t}$, one can make $3m+1$ copies
of $K_{m,t-1}$ and join them at the hubs (namely nodes with largest
degree). As shown in Fig.~\ref{copy}, network $K_{m,t+1}$ may be
obtained by the juxtaposition of $3m+1$ copies of $K_{m,t}$, which
are labeled as $K_{m,t}^{1}$, $K_{m,t}^{2}$, $\ldots$,
$K_{m,t}^{3m}$, and $K_{m,t}^{3m+1}$, respectively.

We continue by exhibiting the procedure of the determination of the
total distance and present the recurrence formula, which allows us
to obtain $D_{\rm tot}(t+1)$ of the $t+1$ generation from $D_{\rm
tot}(t)$ of the $t$ generation. From the obvious self-similar
structure of Koch networks, it is easy to see that the total
distance $D_{\rm tot}(t+1)$ satisfies the recursion relation
\begin{equation}\label{eq:app3}
D_{\rm tot}(t+1)=(3m+1)\,D_{\rm tot}(t) + \Omega_t\,,
\end{equation}
where $\Omega_t$ is the sum over all shortest paths whose endpoints
are not in the same $K_{m,t}^{\theta}$ branch. The solution of
Eq.~(\ref{eq:app3}) is
\begin{equation}\label{eq:app4}
D_{\rm tot}(t) = (3m+1)^{t-1} D_{\rm tot}(1) + \sum_{\tau=1}^{t-1}
(3m+1)^{t-\tau-1} \Omega_\tau\,.
\end{equation}
The paths contributing to $\Omega_t$ must all go through at least
one of the three edge nodes (i.e., grey nodes $X$, $Y$ and $Z$ in
Fig.~\ref{copy}) at which the different $K_{m,t}^{\theta}$ branches
are connected. The analytical expression for $\Omega_t$, called the
length of crossing paths, is found below.

\begin{figure}
\begin{center}
\includegraphics[width=.25\linewidth,trim=120 20 120 20]{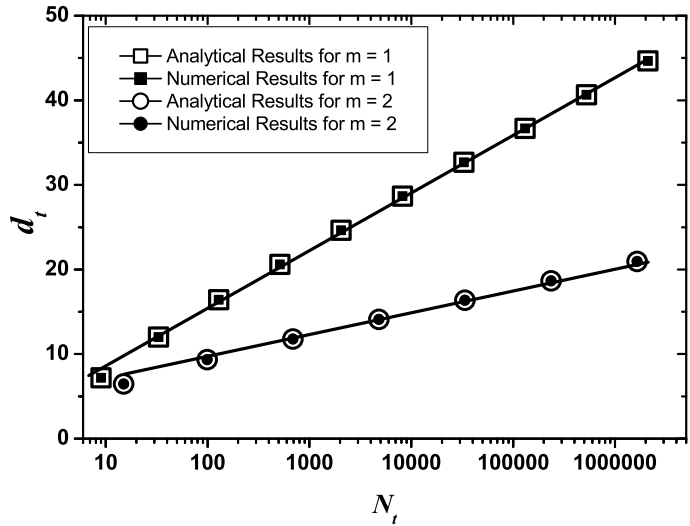}
\end{center}
\caption[kurzform]{\label{distance} Average path length $d_{t}$
versus network order $N_{t}$ on a semi-log scale. The solid lines
are guides to the eyes.}
\end{figure}

Let $\Omega_t^{\alpha,\beta}$ be the sum of the length of all
shortest paths with endpoints in $K_{m,t}^{\alpha}$ and
$K_{m,t}^{\beta}$. According to whether or not two branches are
adjacent, we sort the crossing path length $\Omega_t^{\alpha,\beta}$
into two classes: If $K_{m,t}^{\alpha}$ and $K_{m,t}^{\beta}$ meet
at an edge node, $\Omega_t^{\alpha,\beta}$ rules out the paths where
either endpoint is that shared edge node. For example, each path
contributed to $\Omega_t^{1,2}$ should not end at node $X$. If
$K_{m,t}^{\alpha}$ and $K_{m,t}^{\beta}$ do not meet,
$\Omega_t^{\alpha,\beta}$ excludes the paths where either endpoint
is any edge node. For instance, each path contributed to
$\Omega_t^{2,m+2}$ should not end at nodes $X$ or $Y$. We can easily
compute that the numbers of the two types of crossing paths are
$\frac{3m^2+3m}{2}$ and $3m^2$, respectively. On the other hand, any
two crossing paths belonging to the same class have identical
length. Thus, the total sum $\Omega_t$ is given by
\begin{equation}\label{eq:app6}
\Omega_t = \frac{3m^2+3m}{2}\, \Omega_t^{1,2} + 3m^2\,
\Omega_t^{2,m+2}\,.
\end{equation}
In order to determine $\Omega_t^{1,2}$ and $\Omega_t^{2,m+2}$, we
define
\begin{equation}
s_t = \sum_{i \in K_{m,t},i\ne X}d_{iX}(t)\,. \label{eq:app7}
\end{equation}
Considering the self-similar network structure, we can easily know
that at time $t+1$, the quantity $s_{t+1}$ evolves recursively as
\begin{eqnarray}
s_{t+1}
&=&(m+1)\,s_t+2m\left[s_t+(N_t-1)\right]\nonumber\\
&=&(3m+1)\,s_t+4m\,(3m+1)^{t}.\label{eq:app8}
\end{eqnarray}
Using $s_0=2$, we have
\begin{eqnarray}\label{eq:APL6}
s_t=(4mt+6m+2)\,(3m+1)^{t-1}.
\end{eqnarray}
Have obtained $s_t$, the next step is to compute the quantities
$\Omega_t^{1,2}$ and $\Omega_t^{2,m+2}$ given by
\begin{eqnarray}
  \Omega_t^{1,2} &=& \sum_{\stackrel{i \in K_{m,t}^{1},\,\,j\in
      K_{m,t}^{2}}{ i,j \ne X}} d_{ij}(t+1)\nonumber\\
  &=& \sum_{\stackrel{i \in  K_{m,t}^{1},\,\,j\ \in K_{m,t}^{2}}{ i,j \ne X}} [d_{iX}(t+1) + d_{jX}(t+1)] \nonumber\\
  &=& (N_t-1)\sum_{\stackrel{i \in K_{m,t}^{1}}{ i \ne X}} d_{iX}(t+1) + (N_t-1) \sum_{\stackrel{j \in K_{m,t}^{2}}{ j \ne X}} d_{jX}(t+1) \nonumber\\
  &=& 2(N_t-1)\sum_{\stackrel{i \in K_{m,t}^{1}}{ i \ne X}} d_{iX}(t+1)\nonumber\\
  &=& 2(N_t-1)\,s_t,
\label{eq:app9}
\end{eqnarray}
and
\begin{eqnarray}
  \Omega_t^{2,m+2} &=& \sum_{\stackrel{i \in K_{m,t}^{2},\,i \ne X}{j\in
      K_{m,t}^{m+2},\,j \ne Y}} d_{ij}(t+1)\nonumber\\
  &=& \sum_{\stackrel{i \in K_{m,t}^{2},\,i \ne X}{ j\in
      K_{m,t}^{m+2},\,j \ne Y}} [d_{iX}(t+1) + d_{XY}(t+1)+ d_{jY}(t+1)] \nonumber\\
  &=& 2(N_t-1)\,s_t+(N_t-1)^2\,,
\label{eq:app10}
\end{eqnarray}
where $d_{XY}(t+1)=1$ has been used. Substituting
Eqs.~(\ref{eq:app9}) and (\ref{eq:app10}) into Eq.~(\ref{eq:app6}),
we obtain
\begin{eqnarray}\label{eq:app12}
\Omega_t &=& (9m^2+3m)(N_t-1)\,s_t+3m^2\,(N_t-1)^2\,\nonumber\\
&=&12m(2mt+4m+1)(3m+1)^{2t}.
\end{eqnarray}
Inserting Eqs.~(\ref{eq:app12}) for $\Omega_\tau$ into
Eq.~(\ref{eq:app4}), and using $D_{\rm tot}(1) =48m^2+21m+3$, we can
obtained exactly the expression for $D_{\rm tot}(t)$ as
\begin{equation}\label{eq:app13}
D_{\rm tot}(t) = \frac{(3m+1)^{t-1}}{3} \left
[3m+5+(24mt+24m+4)(3m+1)^{t}\right]\,.
\end{equation}
By plugging Eq.~(\ref{eq:app13}) into Eq.~(\ref{eq:app1}), one can
obtain the analytical expression for $d_t$:
\begin{equation}\label{APL}
d_t = \frac {3m+5+(24mt+24m+4)(3m+1)^{t}}{3(3m+1)\,[2(3m+1)^t+1]},
\end{equation}
which approximates $\frac{4mt}{3m+1}$ in the infinite $t$, implying
that the APL shows a logarithmic scaling with network order. This
again shows that the Koch networks exhibit a small-world behavior.
We have checked our analytic result for $d_t$ given in
Eq.~(\ref{APL}) against numerical calculations for different $m$ and
various $t$. In all the cases we obtain a complete agreement between
our theoretical formula and the results of numerical investigation,
see Fig.~\ref{distance}.

\subsection{Degree correlations}

Degree correlation is a particularly interesting subject in the
field of network
science~\cite{Newman02,Newman03c,MsSn02,PaVaVe01,VapaVe02,ZhZh07},
because it can give rise to some interesting network structure
effects. An interesting quantity related to degree correlations is
the average degree of the nearest neighbors for nodes with degree
$k$, denoted as $k_{\rm nn}(k)$, which is a function of node degree
$k$~\cite{PaVaVe01,VapaVe02}. When $k_{\rm nn}(k)$ increases with
$k$, it means that nodes have a tendency to connect to nodes with a
similar or larger degree. In this case the network is defined as
assortative \cite{Newman02,Newman03c}. In contrast, if $k_{\rm
nn}(k)$ is decreasing with $k$, which implies that nodes of large
degree are likely to have near neighbors with small degree, then the
network is said to be disassortative. If correlations are absent,
$k_{\rm nn}(k)=const$.

We can exactly calculate $k_{\rm nn}(k)$ for Koch networks using
Eqs.~(\ref{ki}) and (\ref{ki2}) to work out how many links are made
at a particular step to nodes with a particular degree. By
construction, we have the following
expression

\begin{eqnarray}\label{Knn1}
k_{\rm nn}(k)&=&{1\over L_v(t_i) k(t_i,t)}\Bigg(
  \sum_{t'_i=0}^{t'_i=t_i-1}m \,L_v(t'_i) k(t'_i,t_i-1)k(t'_i,t) \nonumber\\
 &\quad& +\sum_{t'_i=t_i+1}^{t'_i=t} m\, L_v(t_i) k(t_i,t'_i-1)
 k(t'_i,t)\Bigg)+1
\end{eqnarray}
for $k=2\,(m+1)^{t-t_{i}}$. Here the first sum on the right-hand
side accounts for the links made to nodes with larger degree (i.e.,
$t'_i<t_i$) when the node was generated at $t_i$. The second sum
describes the links made to the current smallest degree nodes at
each step $t'_i>t_i$. The last term 1 accounts for the link
connected to the simultaneously emerging node. In order to compute
Eq.~(\ref{Knn1}), we distinguish two cases according to parameter
$m$: $m=1$ and $m \geq 2$.

When $m=1$, we have
\begin{eqnarray}\label{Knn2}
k_{\rm nn}(k)=t+2.
\end{eqnarray}
Thus, in the case of $m=1$, the networks show absence of
correlations in the full range of $t$. From Eqs.~(\ref{Knn2})
and~(\ref{Nt}) we can easily see that for large $t$, $k_{\rm nn}(k)$
is approximately a logarithmic function of network order $N_t$,
namely, $k_{\rm nn}(k)\sim \ln N_t$, exhibiting a similar behavior
as that of the BA model~\cite{VapaVe02} and the two-dimensional
random Apollonian network~\cite{ZhZh07}.

When $m \geq 2$, equation (\ref{Knn1}) is simplified to
\begin{eqnarray} \label{Knn3}
k_{\rm nn}(k)= \frac{3m+1)}{m-1}\left [\frac{(m+1)^{2}}{3m+1}
\right]^{t_i}- \frac{m+3}{m-1} +\frac{2m}{m+1}(t-t_i).
\end{eqnarray}
Thus after the initial step $k_{\rm nn}(k)$ grows linearly with
time. Writing Eq.~(\ref{Knn3}) in terms of $k$, it is
straightforward to obtain
\begin{eqnarray} \label{knn3}
k_{\rm nn}(k)= \frac{3m+1}{m-1}\,\left [\frac{(m+1)^{2}}{3m+1}
\right ]^{t}\,\left ( \frac{k}{2}\right)^{-\frac{\ln\left
[\frac{(m+1)^{2}}{3m+1}\right ]}{\ln(m+1)}}\nonumber\\
\qquad\qquad\qquad\qquad-\frac{m+3}{m-1}+\frac{2m}{m+1}\,\frac{\ln(\frac{k}{2})}{\ln(m+1)}.
\end{eqnarray}
Therefore, $k_{\rm nn}(k)$ is approximately a power law function of
$k$ with negative exponent, which shows that the networks are
disassortative. Note that $k_{\rm nn}(k)$ of the Internet exhibits a
similar power-law dependence on the degree $k_{\rm nn}(k)\sim
k^{-\omega}$, with $\omega=0.5$ \cite{PaVaVe01}.

\subsection{Spanning trees, spanning forests, and connected spanning
subgraphs}

Spanning trees, spanning forests, and connected spanning subgraphs
are important quantities of networks, and the enumeration of these
interesting quantities in networks is a fundamental
issue~\cite{Wu77,Ly05,BeLyPeSc01,Te05,LiCh83}. However, explicitly
determining the numbers of these quantities in networks is a
theoretical challenge~\cite{TeWa07}. Fortunately, the peculiar
construction of Koch networks makes it possible to derive exactly
the three variables.

\subsubsection{Spanning trees.}

By definition, a spanning tree of any connected network is a minimal
set of edges that connect every node. The problem of spanning trees
is closely related to various aspects of networks, such as
reliability~\cite{Bo86,SzAlKe03}, optimal
synchronization~\cite{NiMo06}, and random walks~\cite{DhDh97}. Thus,
it is of great interest to determine the exact number of spanning
trees~\cite{ZhLiWuZh10}. In what follows we will examine the number
of spanning trees in Koch networks.

Notice that in the Koch networks $K_{m,t}$ there are
$L_\triangle(t)=(3m+1)^t$ triangles, but there are no cycles of
length more than 3. For each of $L_\triangle(t)=(3m+1)^t$ triangles,
to assure that its three nodes are in one tree, two and only two
edges of it must be present. Obviously, there are 3 possibilities
for this. Thus, the total number of spanning trees in $K_{m,t}$,
denoted by $N_{\rm ST}(t)$, is
\begin{eqnarray}\label{STt01}
N_{\rm ST}(t)=3^{L_\triangle(t)}=3^{(3m+1)^t}\,.
\end{eqnarray}

We proceed to represent $N_{\rm ST}(t)$ as a function of the network
order $N_t$, with the aim to provide the relation governing the two
quantities. From Eq.~(\ref{Nt}), we have $(3m+1)^t=\frac{N_t-1}{2}$.
This expression allows one to write $N_{\rm ST}(t)$ in terms of
$N_t$ as
\begin{eqnarray}\label{STt02}
N_{\rm ST}(t)= 3^{(N_t-1)/2}\,.
\end{eqnarray}
Thus, the number of spanning trees in $K_{m,t}$ increases
exponentially with network order $N_t$, which means that there
exists a constant $E_{\rm ST}$, called as the entropy of spanning
trees, describing this exponential growth~\cite{Ly05}:
\begin{equation}\label{STt03}
E_{\rm ST}=\lim_{N_t \rightarrow \infty }\frac{\ln N_{\rm
ST}(t)}{N_t}=\frac{1}{2}\ln 3\,.
\end{equation}

In addition to above analytical computation, according to the
previously known result~\cite{Bi93}, one can also obtain numerically
but exactly the number of spanning trees, $N_{\rm ST}(t)$, by
computing the non-zero eigenvalues of the Laplacian matrix
associated with networks $K_{m,t}$ as
\begin{equation}\label{STt04}
 N_{\rm ST}(t)=\frac{1}{N_t}\prod_{i=1}^{i=N_t-1}\lambda_i(t)\,,
\end{equation}
where $\lambda_i(t)$ ($i = 1, 2,\ldots, N_t-1$) are the $N_t-1$
nonzero eigenvalues of the Laplacian matrix, denoted by
$\textbf{L}_t$, for networks $K_{m,t}$, which is defined as follows:
its non-diagonal element $l_{ij}(t)$ ($i \neq j$) is -1 (or 0) if
nodes $i$ and $j$ are (or not) directly linked to each other, while
the diagonal entry $l_{ii}(t)$ is exactly the degree of node $i$.

Using Eq.~(\ref{STt04}), we have calculated directly the number of
spanning trees in networks $K_{m,t}$, the results from
Eq.~(\ref{STt04}) are fully consistent with those obtained from
Eq.~(\ref{STt01}), showing that our analytical formula is right. It
should be stressed that although the expression of Eq.~(\ref{STt04})
seems compact, it is involved in the computation of eigenvalues of a
matrix of order $N_t \times N_t$, which makes heavy demands on time
and computational resources. Thus, it is not acceptable for large
networks. In particular, by virtue of the eigenvalue method it is
difficult and even impossible to obtain the entropy $E_{\rm ST}$.
Our analytical computation can get around the two difficulties, but
is only applicable to peculiar networks.

\subsubsection{Spanning forests.}

To define spanning forests, we first recall the definition for
spanning subgraph. A spanning subgraph of a network is a subgraph
having the same node as it but having partial or all edges of the
original graph. A spanning forest of a network is a spanning graph
of it that is disjoint union of trees (here an isolated node is
consider as a tree), i.e., a spanning graph without any cycle. The
numeration of number of spanning forests is very interesting since
it corresponds to the partition function of the $q$-state Potts
model~\cite{Wu82} in the limit of $q\rightarrow 0$. For a general
network, it is very hard to count the number of its spanning
subgraphs. But below we will show that for the Koch networks
$K_{m,t}$, the number of spanning subgraphs, $N_{\rm SF}(t)$, can be
obtain explicitly.

Analogous to the enumeration of spanning trees, for each triangle in
$K_{m,t}$, to guarantee the absence of cycle among its three nodes,
at least one edge must be removed. And there are total 7
possibilities for deleting edges of a triangle. Then the number of
spanning forests in $K_{m,t}$ is
\begin{eqnarray}\label{SFt01}
N_{\rm SF}(t)=7^{L_\triangle(t)}=7^{(3m+1)^t}\,,
\end{eqnarray}
which can be rewritten as a function of network order $N_{t}$ as
\begin{eqnarray}\label{SFt01}
N_{\rm SF}(t)=7^{(N_t-1)/2}\,.
\end{eqnarray}
Therefore, $N_{\rm SF}(t)$ also grows exponentially in $N_{t}$,
which allows for defining the entropy of spanning forests of Koch
networks as the limiting value~\cite{BuPe93}
\begin{equation}\label{STt03}
E_{\rm SF}=\lim_{N_t \rightarrow \infty }\frac{\ln N_{\rm
SF}(t)}{N_t}=\frac{1}{2}\ln 7\,.
\end{equation}
Thus, we have obtained the rigorous results for the number of
spanning forests in Koch networks and its entropy.

\subsubsection{Connected spanning subgraphs.}

As the name suggests, a connected spanning subgraph of a connected
network is a spanning subgraph of the network, which remains
connected. By applying a method similar to above, we can compute the
number of connected spanning subgraphs in the Koch networks
$K_{m,t}$, which is denoted by $N_{\rm CSS}(t)$. For any triangle in
$K_{m,t}$, to ensure the connectedness of its three nodes, at most
one edge can be deleted. In other words, two or three edges of it
should be present. On the other hand, for an arbitrary triangle in
$K_{m,t}$, the number of its connected spanning subgraphs is 4.
Then, the number of connected spanning subgraphs is
\begin{eqnarray}\label{CSS01}
N_{\rm CSS}(t)=4^{L_\triangle(t)}=4^{(3m+1)^t}\,,
\end{eqnarray}
which can be further recast in terms of network order $N_{t}$ as
\begin{eqnarray}\label{CSS02}
N_{\rm CSS}(t)=4^{(N_t-1)/2}\,.
\end{eqnarray}
Clearly $N_{\rm CSS}(t)$ increases exponentially in $N_{t}$. Thus,
the entropy of connected spanning subgraphs of Koch networks is
\begin{equation}\label{CSS03}
E_{\rm CSS}=\lim_{N_t \rightarrow \infty }\frac{\ln N_{\rm
CSS}(t)}{N_t}=\ln 2\,.
\end{equation}

\section{Conclusions}

Networks having a complicated enough structure to display nontrivial
properties for real-life systems and statistical models but simple
enough to reveal analytical insights are few and far between. In
this paper, we have presented a mapping that converts the Koch
curves into complex networks, which have many important general
properties observed in real networks. Our networks are characterized
by closed-form, exact formulas for various properties, especially
for numbers for spanning trees, spanning forests, and connected
spanning subgraphs. The rigorous solutions show that the resulting
graphs have a heavy-tailed degree distribution with general exponent
$\gamma \in [2,3]$, logarithmical diameter and average path length
with network order, large clustering coefficient, and degree
correlations. Our analytical technique could guide and shed light on
related studies for deterministic network models by providing a
paradigm for computing the structural features.

In addition to the analytical method and the nontrivial topological
characterizations of our proposed model, another contribution of our
work is mapping the Koch fractals to a family of graphs. With the
the mapping method, a natural bridge between the theory of network
science and the classic fractals has been built. This network
representation technique could find application to some real systems
making it possible to explore the complexity of real-life networked
systems in biological and information fields within the framework of
complex network theory. Recently, a similar recipe has been adopted
for investigating the navigational complexity of
cities~\cite{RoTrMiSn05}, which has also proven useful to the study
of polymer physics~\cite{KaSt05}.

Finally, it should be stressed that for the special case of $m=1$,
the network is completely uncorrelated, it is thus of potential
interest as a null substrate network to corroborate the dynamical
behavior of complex systems, since the analytic solution of
dynamical processes is usually available only for uncorrelated
networks~\cite{CaNeStWa00,CoErAvHa00,CoErAvHa01,PaVe01a}, the
generation algorithm of which is often more difficult than one would
expect \emph{a priori}~\cite{CaBoPa05}.

Before closing the paper, it should be mentioned that in a previous
paper Barab\'asi Ravasz and Vicsek proposed a deterministic model,
hereafter called the BRV model~\cite{BaRaVi01}, which is the
progenitor of deterministic network models. Although the BRV model
is also scale-free and small-world~\cite{ZhLiGaZhGu09}, its degree
exponent $\gamma$ is a constant $1+\frac{\ln 3}{\ln 2}$, and its
clustering coefficient is zero because there is no triangle in the
BRV model. In this context, our model addressed here can mimic real
systems better than the BRV model.

\subsection*{Acknowledgment}

We would like to thank Yichao Zhang for his assistance. This
research was supported by the National Natural Science Foundation of
China under Grants No. 60704044, No. 60873040, and No. 60873070, the
National Basic Research Program of China under Grant No.
2007CB310806, and Shanghai Leading Academic Discipline Project No.
B114. Shuyang Gao also acknowledges the support by Fudan's
Undergraduate Research Opportunities Program, and Hongjuan Zhang
acknowledges the support by Shanghai Key Laboratory of Intelligent
Information Processing, China. Grant No. IIPL-09-017.


\end{document}